\documentstyle[preprint,prl,aps,amssymb,graphicx]{revtex}
\input{epsf}

\begin{document}

\title{Quantum dynamical properties of a two-photon 
non-linear Jaynes-Cummings model}
\author{Guido Berlin and J. Aliaga} 
\address {Departamento de F\'{\i}sica, Facultad de Ciencias Exactas
y Naturales,\\ 
Universidad de Buenos Aires, (1428) Buenos Aires, Argentina}
\date{\today}

\maketitle

\begin{abstract}
In this paper we study a two-photon 
time-dependent Jaynes-Cummings model 
interacting with a Kerr-like medium. 
We assumed that the electromagnetic field is in
different states such as coherent, squeezed vacuum and pair coherent, and
that the atom is initially in the excited state.
We studied the temporal evolution of the population of the excited level,
and the second order coherence function.
The results 
obtained show that this system has some similarities 
with the two-mode Stark system. 
We analize two photon entanglement for
different initial conditions.

\end{abstract}
{{\bf PACS numbers}: 42.50.-p, 42.50.Dv.\\ 
{\bf Corresponding author}: Jorge Aliaga.  e-mail:  jaliaga@df.uba.ar\\
\address: Departamento de F\'{\i}sica, Facultad de Ciencias Exactas
y Naturales,\\ 
Universidad de Buenos Aires, Pabell\'on I, (1428) Buenos Aires, Argentina\\
FAX: +54 - 11 - 4576 - 3357}

\section{Introduction}

One of the most fundamental models in quantum optics is the
interaction of a single two-level atom with a single quantized mode of
radiation, described by the Jaynes-Cummings Hamiltonian \cite
{Jaynes63,Shore93}. Despite being simple enough to be analytically soluble
in the rotating-wave approximation, this model is able to describe the 
quantum-mechanical aspects of the interaction between light and
matter. It has led to nontrivial predictions, such as the existence of
``collapses" and ``revivals" in the atomic excitation \cite{Eberly80}, and
has also allowed a deeper understanding of the dynamical entangling and
disentangling of the atom-field system in the course of time \cite
{Phoenix88}. The interest in the Jaynes-Cummings model (JCM) 
is mainly due to the fact that some of its predictions are nowadays 
accessible to experimental verification \cite{Exper}. 
A JCM interaction can be experimentally realized in cavity-QED
setups and, as an effective interaction, in
laser-cooled trapped ions.

During the last decade many theoretical and experimental 
efforts have been done in order to study two-photon 
processes involving atoms inside a cavity, stimulated by the
experimental realization of a two-photon cascade micromaser \cite{Brune} 
Two-photon processes are also an efficient way of generating squeezed 
states of the electromagnetic field. It has also been  
established that two-photon degenerate atoms inside a 
Kerr-like medium can generate squeezing amplification \cite{Squeez_Amp}. 
The recent discovery of new materials with very high Kerr coupling
\cite{H_Kerr} opened the possibility for the implementation of new
experiments that generates entangled states \cite{Vitali} 
that can be used for a perfect Bell-state discrimination. 

A system composed by a three level atom in $\Xi$, $\Lambda$, and $V$ 
configurations interacting with two modes of the electromagnetic field was 
proposed and studied more than ten years ago by Yoo and 
Eberly \cite{Yoo}. Following these guidelines, Gou \cite{Gou1} 
investigated the $\Xi$ configuration
when the intermediate level can be adiabatically eliminated \cite{Alsing}.
This approximation turns the original bilinear photon-level interaction into a 
three-linear one (usually called ``non-linear non-degenerate two-photon'' interaction).
This model has been broadly used in order to study the time evolution of the
atomic and photon operators, the second order coherence function, the one and 
two-mode squeezing, the atomic-dipole squeezing and the emission spectra
\cite{Gou1,Joshi,Gerry1,Abdel,Ashraf1,Ashraf2,Iwasawa}.
Usually, the two-level system has been considered initially in the excited state 
and the two-photons have been chosen initially in
two independent coherent states \cite{Gou1,Abdel,Ashraf1,Iwasawa}, 
two mode squeezed states \cite{Gou1}, pair-coherent states \cite{Joshi}, 
correlated SU(1,1) coherent states \cite{Gerry1}, etc.
The $\Lambda$ configuration, also called the Raman coupled model, 
when the intermediate level can be adiabatically eliminated was
studied by Abdalla, Ahmed and Obada \cite{Abdalla}, and independently
by Gerry and Eberly \cite{Gerry2}. There have been investigations of the atomic 
inversion, the appearance of antibunched light, the violations of the 
Cauchy-Schwartz inequality, population trapping, and squeezing
\cite{Abdalla,Gerry2,Cardimona,Gerry3,Deb1}. 
Some similarities in the Rabi frequencies of both configurations for 
some special conditions of the parameters were reported \cite{Gerry1}, 
and the connection between these configurations was deeply studied in
Ref. \cite{JC2}. It was reported that 
for the $\Xi$ \cite{Gou2,Souza} and $\Lambda$ 
\cite{Nasreen,Li} models, when doing the adiabatic approximation, the 
appearance of Stark shifts must be taken into account. In
Ref. \cite{JC2} the Stark shifts were neglected in order to separate their
contribution to the nontrivial dynamics studied. 
The
intensity-dependent Stark shifts caused by off-resonant levels \cite{Joshi2} 
were modeled by Moya-Cessa, Bu$\check{{\rm z}}$ek 
and Knight \cite{Moya2}
using a JCM with an intensity-dependent shift of the two-level transition.
Recently, some similarities and differences between the models used in
order to describe a cavity
filled with a Kerr-like medium, modeled by an anharmonic oscillator,
(first analyzed in detail by Jex and Bu$\check{{\rm z}}$ek \cite{Buzek} 
and discussed in many articles thereafter \cite{Werner,Ho,Joshi2,Gruver2}) 
and Stark effects have been studied \cite{Moya,Kerr-Stark}.

In this paper we study a two-photon time-dependent 
Jaynes-Cummings model interacting with a nonlinearity that can
be, for instance, a Kerr-like medium. 
The problem is solved using a technique based on obtaining those
observables dynamically related \cite{JC2}, and then solving their
temporal evolution.  We assumed that the electromagnetic field is in
different states such as coherent, squeezed vacuum and pair coherent, and
that the atom is initially in the excited state.
We studied the temporal evolution of the population of the excited level,
and the second order coherence function.
The results 
obtained show that this system has some similarities 
with the two-mode Stark system. 

\section{The system}

\noindent We study an effective two-level atom \cite{JC2}, whose levels
 $|g>$ and $|e>$,
with energies $E_1$ and $E_2$, interact  with  two modes of
electromagnetic radiation of frequencies  $\omega_1$ and $\omega_2$
inside a non-linear Kerr-like medium.
In two-photon processes there are more than two levels involved,
but it is possible to neglect them if we assume that 
$\omega_1+\omega_2 \approx E_1 - E_2$  ($\hbar=1$) and we consider that the
transition frequencies between 
$|e>$, $|g>$ and the intermediate levels are different from the 
frequencies of the field.
 So, the JCM in the rotating wave
approximation, ($\Xi$ configuration), reads \cite{JC2}:
\begin{eqnarray}
\label{ham2m}
\hat H = \sum_{i=1}^{2} E_{i} \hat b_{i}^{\dagger} \hat b_{i} +
\sum_{i=1}^{2} \omega_{i} \hat a_{i}^{\dagger} \hat a_{i} +
T(t)(\gamma \hat a_{1} \hat a_{2} \hat b_{1} \hat b_{2}^{\dagger} +
\gamma^{*} \hat b_{2} \hat b_{1}^{\dagger} \hat a_{2}^{\dagger} \hat
a_{1}^{\dagger})\; , 
\end{eqnarray}
where $a_{i}$, $a_{i}^{\dagger}$, $b_{i}^{\dagger}$ and $b_{i}$, $i = 1,2$, 
are creation and annihilation bosonic and fermionic operators, respectively,
and $\gamma$ is the coupling constant between the atomic levels and the fields.

The Kerr medium can be modeled by an anharmonic oscillator coupled to
the field \cite{Buzek,Werner,Ho,Joshi2,Gruver2,Imoto}. Using the adiabatic 
approximation
($\omega_{medium}\ll \omega_{field}$) the non-linear medium can be
represented by a non-linear power of the field. 
Thus, the total Hamiltonian reads:
\begin{eqnarray}
\label{hamkerr}
\hat{H} &=& {\sum^2_{i=1}} E_i \hat{b}^{\dagger}_i \hat{b}_i 
+ {\sum_{i=1}^2} \omega_i \hat{a}^{\dagger}_i \hat{a}_i 
+ T(t)(\gamma \hat{a}_1 \hat{a}_2 \hat{b}_1 \hat{b}^{\dagger}_2 
+ \gamma^{\ast} \hat{b}_2 \hat{b}^{\dagger}_1 
\hat{a}^{\dagger}_1 \hat{a}^{\dagger}_2)\nonumber\\
&& + \chi_1 (\hat{a}^{\dagger}_1)^2 (\hat{a}_1)^2 + 
\chi_2 (\hat{a}^{\dagger}_2)^2 (\hat{a}_2)^2 
+2{\sqrt{\chi_1 \chi_2}}\hat{a}^{\dagger}_2 \hat{a}^{\dagger}_1
 \hat{a}_1  \hat{a}_2 
\end{eqnarray}
It is important to notice that three non-linear terms appear due to
the presence of the non-linear Kerr media. Two of them are similar
to the ones appearing in the case of one mode while the third one 
is a bilinear connection between the modes. This bilinear interaction
is the one proposed in Refs. \cite{H_Kerr,Vitali} as a way of generating
cat's states.

The dynamical evolution of the mean values of the operators can be obtained by
using the Ehrenfest equation,
\begin{equation}
{d \langle \hat O_{j}\rangle_{t} \over{d t}}
= -\sum_{i} g_{ij} \langle \hat O_{i} \rangle\; , 
\end{equation}
where $g_{ij}$ are the structure constants of a semi-Lie Algebra
closed under commutation with the Hamiltonian \cite{JC2}, i.e.
\begin{equation}
\left[ \hat{H}\left( t \right),
\hat{O}_i \right] = 
i \hbar \sum_i g_{ji} \left( t \right)
\hat{O}_j\; .\nonumber
\end{equation}
The operators defined via the previous equation are called
{\it relevant operators} (RO)\cite{Gruver3}.
The relevant operators for the two-modes Jaynes-Cummings Hamiltonian
read
\begin{mathletters}
\label{RO}
\begin{eqnarray}
\hat{N}^{n,m}_1 &\equiv& (\hat{a}^{\dagger}_1)^n (\hat{a}^{\dagger}_2)^m
\hat{b}^{\dagger}_1 \hat{b}_1 (\hat{a}_2)^m (\hat{a}_1)^n \; \\
\hat{N}^{n,m}_2 &\equiv& (\hat{a}^{\dagger}_1)^n (\hat{a}^{\dagger}_2)^m
\hat{b}^{\dagger}_2 \hat{b}_2 (\hat{a}_2)^m (\hat{a}_1)^n \; ,\\
\hat{\Delta}^{n,m}_1 &\equiv& (\hat{a}^{\dagger}_1)^n
(\hat{a}^{\dagger}_2)^m \hat{a}^{\dagger}_1 \hat{a}_1 
(\hat{a}_2)^m (\hat{a}_1)^n \; , \\
\hat{\Delta}^{n,m}_2 &\equiv& (\hat{a}^{\dagger}_1)^n
(\hat{a}^{\dagger}_2)^m \hat{a}^{\dagger}_2 \hat{a}_2 (\hat{a}_2)^m (\hat{a}_1)^n \;
,\\
\hat{I}^{n,m} &\equiv& (\hat{a}^{\dagger}_1)^n (\hat{a}^{\dagger}_2)^m
(\gamma \hat{a}_1 \hat{a}_2 
\hat{b}_1 \hat{b}^{\dagger}_2 + \gamma^{\ast}
 \hat{a}^{\dagger}_1 \hat{a}^{\dagger}_2 
\hat{b}^{\dagger}_1 \hat{b}_2) (\hat{a}_2)^m (\hat{a}_1)^n \; ,\\
\hat{F}^{n,m} &\equiv& (\hat{a}^{\dagger}_1)^n (\hat{a}^{\dagger}_2)^m
i(\gamma \hat{a}_1 \hat{a}_2
 \hat{b}_1 \hat{b}^{\dagger}_2 - 
\gamma^{\ast} \hat{a}^{\dagger}_1 
\hat{a}^{\dagger}_2 \hat{b}^{\dagger}_1 \hat{b}_2) 
(\hat{a}_2)^m (\hat{a}_1)^n\; ,\\
\hat{N}^{n,m}_{2,1} &\equiv& (\hat{a}^{\dagger}_1)^n
(\hat{a}^{\dagger}_2)^m 
\hat{b}^{\dagger}_2 
\hat{b}_2 \hat{b}^{\dagger}_1 \hat{b}_1
 (\hat{a}_2)^m (\hat{a}_1)^n \; .
\end{eqnarray}
\end{mathletters}
Notice that $\hat{I}^{n,m}$ and $\hat{F}^{n,m}$ are the operators that
have the information about the correlation between the modes and the
atom levels (i.e. the entanglement).

Using the Ehrenfest theorem, the evolution equations of the relevant
operators are
\begin{mathletters}
\label{Set1}
\begin{eqnarray}
{d \left< \hat{N}^{n,m}_1 \right> \over dt} &=& T(t) \hat{F}^{n,m} + nT(t)
\hat{F}^{n-1,m} + mT(t) \hat{F}^{n,m-1} \nonumber\\
&&+ nmT(t) \hat{F}^{n-1,m-1}\; ,\\
{d \left< \hat{N}^{n,m}_2 \right> \over dt} &=& -T(t) \hat{F}^{n,m}\; ,\\
{d \left< \hat{\Delta}^{n,m}_1 \right> \over dt} &=&
(n+1)T(t) \hat{F}^{n,m} + mT(t) \hat{F}^{n,m-1} \nonumber\\
&&+ mT(t) \hat{F}^{n+1,m-1} + nmT(t) \hat{F}^{n,m-1} \; ,\\
{d \left< \hat{\Delta}^{n,m}_2 \right> \over dt} &=&
(m+1)T(t) \hat{F}^{n,m} + nT(t) \hat{F}^{n-1,m} \nonumber\\
&&+ nT(t) \hat{F}^{n-1,m+1}+ nmT(t) \hat{F}^{n-1,m}\; ,\\
{d \left< \hat{N}^{n,m}_{2,1} \right> \over dt} &=& 0\; ,\\
{d \left< \hat{I}^{n,m} \right> \over dt} &=&
\left[ \alpha
- 2 \left(n \chi_1 + m\chi_2 +(m+n+1){\sqrt{\chi_1 \chi_2}}
 \right) \right]  \hat{F}^{n,m} \nonumber\\
&&-2 \left( \chi_1 + {\sqrt{\chi_1 \chi_2}} \right)
  \hat{F}^{n+1,m} -2 \left( \chi_2 + {\sqrt{\chi_1 \chi_2}}\right)
 \hat{F}^{n,m+1}\; ,\\
{d \left< \hat{F}^{n,m} \right> \over dt} &=&
-\left[ \alpha
- 2 \left( n \chi_1 + m\chi_2 + \left(m+n+1 \right){\sqrt{\chi_1 \chi_2}}
\right) \right]  \hat{I}^{n,m} \nonumber\\
&&+2 \left( \chi_1 + {\sqrt{\chi_1 \chi_2}} \right)
 \hat{I}^{n+1,m} + 2 \left( \chi_2 +{\sqrt{\chi_1 \chi_2}} \right)
  \hat{I}^{n,m+1} \nonumber\\
&&+2|\gamma|^2 T(t)[ (n+1)(m+1) \hat{N}^{n,m}_2 
- (n+1)(m+1) \hat{N}^{n,m}_{2,1} \nonumber\\
&&+ (n+1) \hat{N}^{n,m+1}_2-(n+1) \hat{N}^{n,m+1}_{2,1}
+(m+1) \hat{N}^{n+1,m}_2 \nonumber\\
&&- (m+1) \hat{N}^{n+1,m}_{2,1}- \hat{N}^{n+1,m+1}_1 
+\hat{N}^{n+1,m+1}_2 ]\; .
\end{eqnarray}
\end{mathletters}
where $\alpha = E_2 - E_1 -\omega_1  - \omega_2$.  

So, using Eqs. (\ref{Set1}) it is possible to evaluate the temporal evolution of
the RO. These operators close a Lie Algebra.
Thus, any extension of the two-photon JCM
built up adding terms proportional to RO will have similar evolution equations.
This fact will be used in the following section in order to study the two-photon
JCM with Stark shifts.

Equations (\ref{Set1}) can be numerically solved for any temporal dependence
of the atomic-field interaction, $T(t)$.
For the time-independent case, Eqs. (\ref{Set1}) can be analytically solved using 
the series expansion in terms of commutators with the Hamiltonian \cite{JC2,Gruver2,Gruver3}
\begin{eqnarray}
\label{sum}
\left< \hat{O} \right>_t = \left< \hat{O} \right>_0
 + \sum_{n\geq 0} {1 \over {n!}}
 \left({t \over {i \hbar}}\right)^n \left< \left[ \dots \left[\hat{O},\hat{H}\right],
\dots \dots,\hat{H} \right] \right>_0\; .
\end{eqnarray}
From Eqs. (\ref{Set1}) we can notice that the evolution equations for all the 
operators different from $\hat{F}^{n,m}$ depend on $\hat{F}^{n,m}$.
So, the double commutator $\left[\hat{H},\left[\hat{H},\hat{F}^{n,m}\right] \right]$
plays a central role. This double commutator reads
\begin{eqnarray}
\label{dobconm}
\left[\hat{H},\left[\hat{H},\hat{F}^{n,m}\right] \right] &=& 
\beta^2_{n,m}\hat{F}^{n,m} + \left(\beta^2_{n+1,m} - \beta^2_{n,m} \right)
\hat{F}^{n+1,m}+\nonumber \\
&&  \left(\beta^2_{n,m+1} - \beta^2_{n,m} \right)\hat{F}^{n,m+1}
 + \left[ 8 \epsilon_1 \epsilon_2 + 4 |\gamma|^2 \right]
 \hat{F}^{n+1,m+1}  +\nonumber \\ 
&&4 \epsilon^2_1 \hat{F}^{n+2,m} + 4 \epsilon^2_2 \hat{F}^{n,m+2}\; ,
\end{eqnarray}
where the generalized Rabi frequency 
$ \beta^2_{n,m}=\left[\alpha - 2\left(n \chi_1 + 
m \chi_2 + (n+m+1) {\sqrt{\chi_1 \chi_2}}\right)
\right]^2 + 4 |\gamma|^2
\left(n+1\right) \left(m+1 \right)$ and 
$\epsilon_i=\chi_i + {\sqrt{\chi_1 \chi_2}}$. 
It is important to notice that the term  proportional to
${\sqrt{\chi_1 \chi_2}}$
appearing in $\beta^2_{n,m}$ and $\epsilon_i$
is due to the bilinear connection between modes.
Equation (\ref{dobconm}) generates {\it paths}, or different ways
of dynamically connecting the RO, representing quantum correlations
\cite{JC2,Gruver2,Kerr-Stark,Gruver3}.
From Eq. (\ref{dobconm}) it can be seen that those terms proportional to
$\epsilon_i$ are due to the presence of the Kerr medium. 
Notice that taking into account the bilinear connection between modes
doubles the value of $\epsilon_i$ in the $\chi_1 = \chi_2$ case usually 
studied.

Making use of Eq. (\ref{dobconm}) we can obtain the temporal 
evolution of the population of the excited level
($\langle \hat N_{2}^{0,0} \rangle_{t}$)
 and
the second order intermodes coherence function
(defined as $g^2_{12}(t)={{\langle  \hat{a}^\dagger_1  \hat{a}_1
\hat{a}^\dagger_2  \hat{a}_2 \rangle}
\over {\langle \hat{a}^\dagger_1 \hat{a}_1 \rangle
\langle \hat{a}^\dagger_2 \hat{a}_2 \rangle } } -1$), 
for different states of the field, (See appendix).

This is depicted in Figs.\ \ref{fig1}-\ref{fig3}.  In all cases, 
we assume the atom initially in the excited state 
($\langle \hat N_{1}^{0,0} \rangle(0)=0$,
$\langle \hat N_{2}^{0,0} \rangle(0)=1$).
The phenomenological third-oder nonlinear susceptibility for the 
Kerr media $\chi=\chi_1=\chi_2$ takes the values $0$, $0.5$ and $1$,
which, in principle, could be obtained in the ultra-high susceptibility Kerr cells 
recently studied \cite{H_Kerr}. We also assume that there are initially
10 photons in each mode ($|\alpha_i|^2=10)$.
We analyze the influence of the non-linear medium in the
temporal evolution of the excited level and we find 
inhibited decay as it can be seen from Figs.\ 
\ref{fig1}-b,\ref{fig2}-b,\ref{fig3}-b.
In the Pair Coherent state case (Fig.\ \ref{fig3}-b), we find that the
revivals are of such a kind that the atom recovers its initial 
population in each revival, and this effect is increased as 
we make more important the coupling with the non-linear medium.
Another interesting feature is that in the squeezed vacuum
state (Fig.\ \ref{fig2}-b) we see that the revivals are regular and sharp, 
and become more periodic as we increase the strength of the coupling
with the Kerr medium.
We can see from Figs.\ \ref{fig1}-a and \ref{fig3}-a, that 
the two-photon JCM presents
antibunching, and that this effect is attenuated by a 
stronger interaction with the non-linear medium.  In the particular
case of the squeezed vacuum, we notice that the field recovers its 
initial intermodes coherence when the 
excited level population
reaches a maximum.

Finally, we want to mention that $\Lambda$ configuration can be studied
by using the canonical transformation $\hat{\tilde{a}}_1 = \hat{a}^\dagger_1$
and redefining the RO following Ref. \cite{JC2}.

\section{Stark effect}

The Stark effect is related to changes in the atomic
energy levels due to virtual transitions from levels out of resonance.
The effective Hamiltonian taking into account the Stark effect reads
 \begin{eqnarray}
\label{hamstark2m}
\hat H &=& \sum_{i=1}^{2} E_{i} \hat b_{i}^{\dagger} \hat b_{i} +
\sum_{i=1}^{2} \omega_{s_i} \hat a_{i}^{\dagger} \hat a_{i} +
T(t)(\gamma \hat a_{1} \hat a_{2} \hat b_{1} \hat b_{2}^{\dagger} +
\gamma^{*} \hat b_{2} \hat b_{1}^{\dagger} \hat a_{2}^{\dagger}
\hat a_{1}^{\dagger})\nonumber\\
&& + \hat a_{1}^{\dagger} \hat a_{1}\left(\eta_1
\hat b_{1}^{\dagger} \hat b_{1}+ \eta_2
\hat b_{2}^{\dagger} \hat b_{2} \right)+
\hat a_{2}^{\dagger} \hat a_{2}\left(\eta_1
\hat b_{1}^{\dagger} \hat b_{1}+ \eta_2
\hat b_{2}^{\dagger} \hat b_{2} \right)\; .
\end{eqnarray}
Both the Kerr Hamiltonian and the Stark Hamiltonian
can be written in terms of RO (for the one-mode case see Ref. 
\cite{Kerr-Stark}. Using the same technique is is easy to see that if
\begin{eqnarray}
\omega_1 + \sqrt{\chi_1\chi_2} &=& \omega_{s_1} + \eta_1  \nonumber \\ 
\omega_2 + \sqrt{\chi_1\chi_2} &=& \omega_{s_2} + \eta_2   \nonumber \\
\eta_1 - \eta_2 &=& 2 \epsilon_1 =  2 \epsilon_2 \; ,
\end{eqnarray}
the evolution equations for the relevant operators
will be the same. In this case, for the RO here studied, both systems 
will have the same temporal evolution \cite{Kerr-Stark}. This will not 
be the case if we study the temporal evolution of other R.O. like the
electromagnetic field $(\hat{a} + \hat{a}^{\dagger})$.

\section{Conclusion}

In the present paper we analyzed the non-degenerate two-photon
JCM in the presence of a Kerr media.
We have identified those relevant operators that are interrelated with
the dynamics of the system.
We find that the dynamical evolution of the relevant operators
for this problem and for the model that takes into account Stark 
shifts are the same for these RO and for some values of the constants.
We analyze the influence of the non-linear medium in the
temporal evolution of the excited level and we find 
inhibited decay.
In the Pair Coherent state case, we observe that the
revivals are of such a kind that the modes of the field
are disentangled, and this effect is increased as 
we make more important the coupling with the non-linear medium.
We also notice that in the squeezed vacuum
state the revivals are regular and sharp, and that
the modes of the field are entangled.
We also see that this system presents
antibunching, and that this effect is attenuated by a 
stronger interaction with the non-linear medium.  
Thus, this system can be used in order to obtain
very different two photon entanglement situations, depending on
the initial state of the field that is considered. This fact
can be very useful in experimental realizations.

\section*{acknowledgments}

The authors acknowledge support from Consejo Nacional de Investigaciones
Cient\'{\i}ficas y T\'ecnicas (Conicet), Argentina.

\appendix
\section*{Exact expressions for the time evolution of the
Relevant Operators}

In the time independent case, making use of equations (\ref{dobconm},\ref{sum})
we can obtain an exact expression for all the Relevant Operators.  For 
instance
$\langle \hat N_{1}^{0,0} \rangle(t)$
reads:

\begin{eqnarray}
\label{n1dete}
\langle \hat N_{1}^{0,0} \rangle(t)=\langle
\hat N_{1}^{0,0} \rangle(0)  &-&
\sum_{j,k=0}^{\infty}
\frac{S_{j,k}}{\beta_{j,k}}
\sum_{n=j}^{\infty} \sum_{m=k}^{\infty} a^{n,m}_{j,k}
\langle \hat F^{n,m} \rangle_{0} \nonumber\\
&-&
\sum_{j,k=0}^{\infty}
\frac{C_{j,k}}{\beta_{j,k}^{2}}
\sum_{n=j}^{\infty} \sum_{m=k}^{\infty} a^{n,m}_{j,k}
[ 2 \epsilon_1 + 2 \epsilon_2
- [  \alpha - 2(n \epsilon_1 + m \epsilon_2 \nonumber \\
 &+& \sqrt{\chi_1 \chi_2}) ]]
\langle \hat I^{n,m} \rangle_{0} + 2 |\gamma|^2
[\langle \hat N_{2}^{n+1,m+1} \rangle -  
\langle \hat N_{1}^{n+1,m+1} \rangle] \nonumber \\
&+& 2 |\gamma|^2 (n+1)(m+1)[\langle \hat N_{2}^{n,m} \rangle
- \langle \hat N_{2,1}^{n,m} \rangle] \nonumber \\
&+& 2 |\gamma|^2 (n+1)[\langle \hat N_{2}^{n,m+1} \rangle
- \langle \hat N_{2,1}^{n,m+1} \rangle]\; ,
\end{eqnarray}

\noindent where $S_{j,k}=sin(\beta_{j,k}t)$, $C_{j,k}
=cos(\beta_{j,k}t) - 1$, and $a^{n,m}_{j,k} = (-1)^{n+m+j+k+1}/
[(n-j)! j! (m-k)! k!]$.\\
\noindent We can evaluate this expression for different initial 
conditions for the population of the atomic levels, and the
state of the field.  

If we assume that the field is initially in a coherent state,

\begin{eqnarray}
|\alpha_{1} \alpha_{2}>=\sum_{i=0}^{\infty}P_{n_1}P_{n_2}|n_1
n_2>\; ,
\end{eqnarray}
where
\begin{eqnarray}
P_{n_{i}}={{e^{|\alpha|^2}{|\alpha|^2}^{n_{i}}}\over{n_{i}!}}\; ,
\end{eqnarray}
\noindent then the expression (\ref{n1dete}) takes the form:
\begin{equation}
\label{N1coherente}
\langle \hat N_{1}^{0,0} \rangle_{t} = 1 +
2|\gamma|^{2}
\sum_{j,k=0}^{\infty} \frac{1}{j! k!}
\frac{C_{j,k}}{\beta_{j,k}^{2}}
\langle \hat \Delta_{1}^{0,0} \rangle_{0}^{j}
\langle \hat \Delta_{2}^{0,0} \rangle_{0}^{k}
e^{-\langle \hat \Delta_{1}^{0,0} \rangle_{0}}
e^{-\langle \hat \Delta_{2}^{0,0} \rangle_{0}}\; ,
\end{equation}
\noindent where $\Delta_{i}^{0,0}$ ($i=1,2$) is the number of
photons present in each mode of the field.

The same treatment can be applied to the case of having 
the field in the Squeezed Vacuum state \cite{walls},
which can be written
in the following way,

\begin{eqnarray}
\label{sv}
|r, \phi>=(cosh (r))^{-1} \sum_n (e^{i\phi}tanh (r))^{n} |n, n>\; .
\end{eqnarray}

\noindent In this case the expression (\ref{n1dete}) reads,

\begin{eqnarray}
\label{n1sqv}
\langle \hat N_{1}^{0,0} \rangle_{t} = 1 +
\frac{2|\gamma|^{2}}{cosh^2(r)}
\sum_{k=0}^{\infty} (k+1)^2
tanh^2(r) 
\frac{C_{k,k}}{\beta^2_{k,k}} \; .
\end{eqnarray}

Finally, we assume the field is in the Pair Coherent state
\cite{agarwal},

\begin{equation}
|\xi,q>=N_q \sum_{n=0}^{\infty}{{\xi^n}\over{[n!(n+q)!)]^{1 \over 2}}}
|n+q,n>\;,
\end{equation}

\noindent where $N_q$ is a normalization constant.  In this case we have:

\begin{eqnarray}
\label{n1pcs}
\langle \hat N_{1}^{0,0} \rangle_{t} = 1 + 2|\gamma|^2 N_q 
\sum_{k=0}^{\infty}(k+1)(k+q+1) 
\frac{|\xi|^{2k}}{k!(k+q)!} C_{k,k}
\end{eqnarray}

%
%

\begin{center}
\begin{figure}
\includegraphics[width=18cm, height=16cm ]{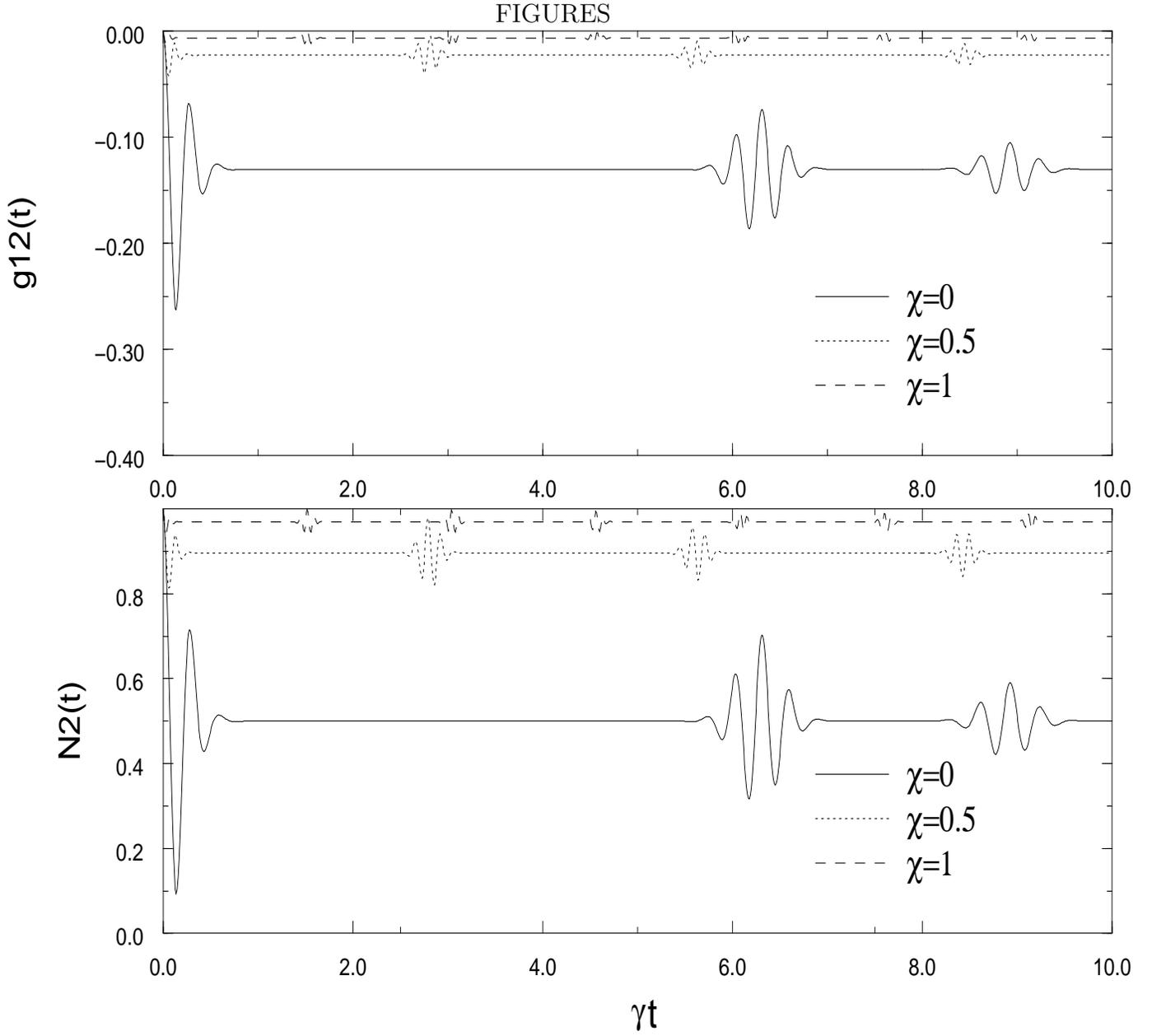}
\caption{Figure 1.  $\langle \hat N_{2}^{0,0} \rangle(t)$
and $g^2_{12}(t)$ for
coherent state field, $\langle \hat N_{2}^{0,0} \rangle(0)=1$,
$\langle \hat N_{1}^{0,0} \rangle(0)=0$
and $|\alpha_i|^2=10)$.}
\label{fig1}
\end{figure}
\end{center}

\newpage
\begin{center}
\begin{figure} 
\includegraphics[width=18cm, height=16cm ]{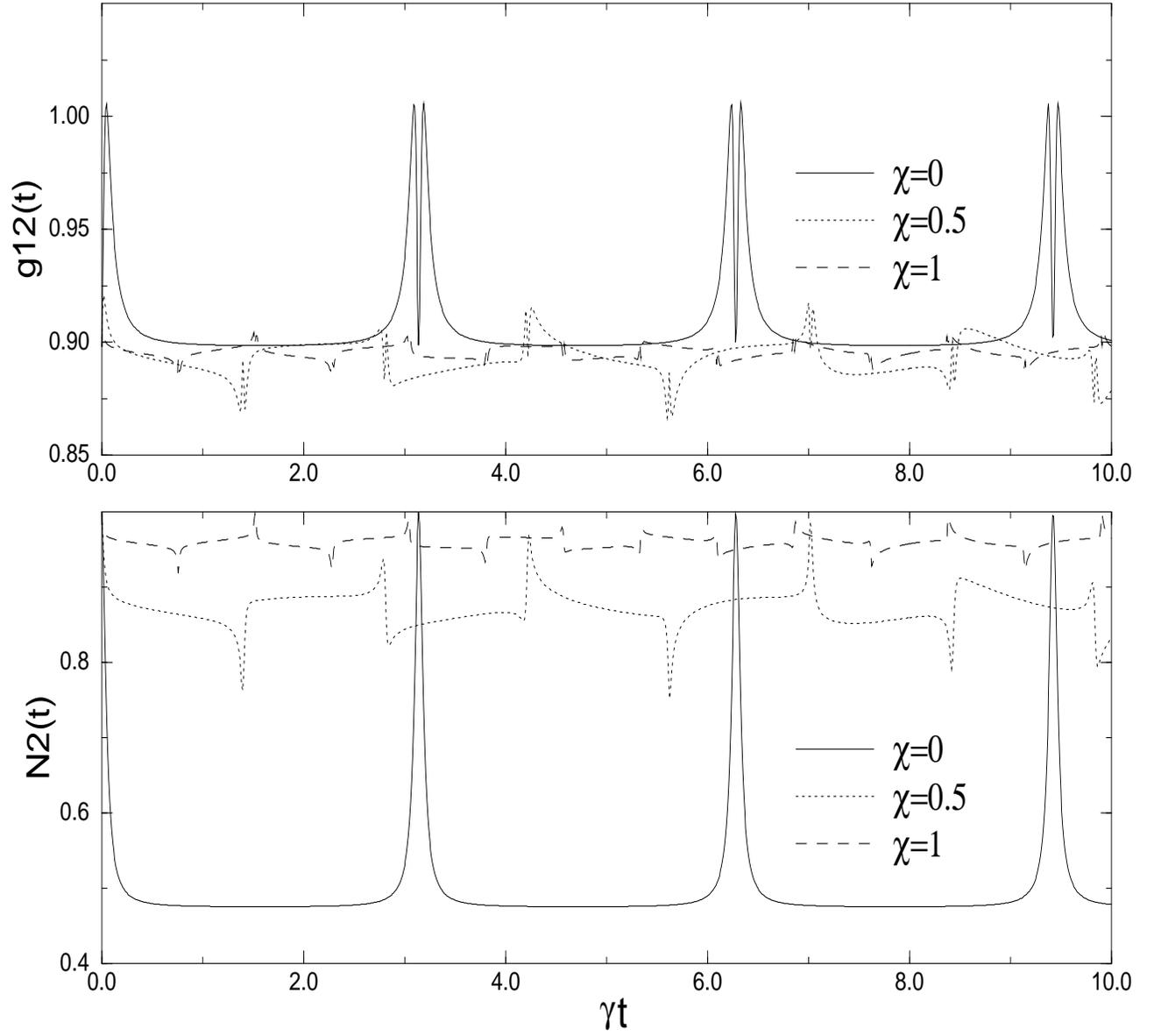}
\caption{Figure 2.   Same as figure 1 but field
in squeezed vacuum.}
\label{fig2}                      
\end{figure}
\end{center}

\begin{center}
\begin{figure}
\includegraphics[width=18cm, height=16cm ]{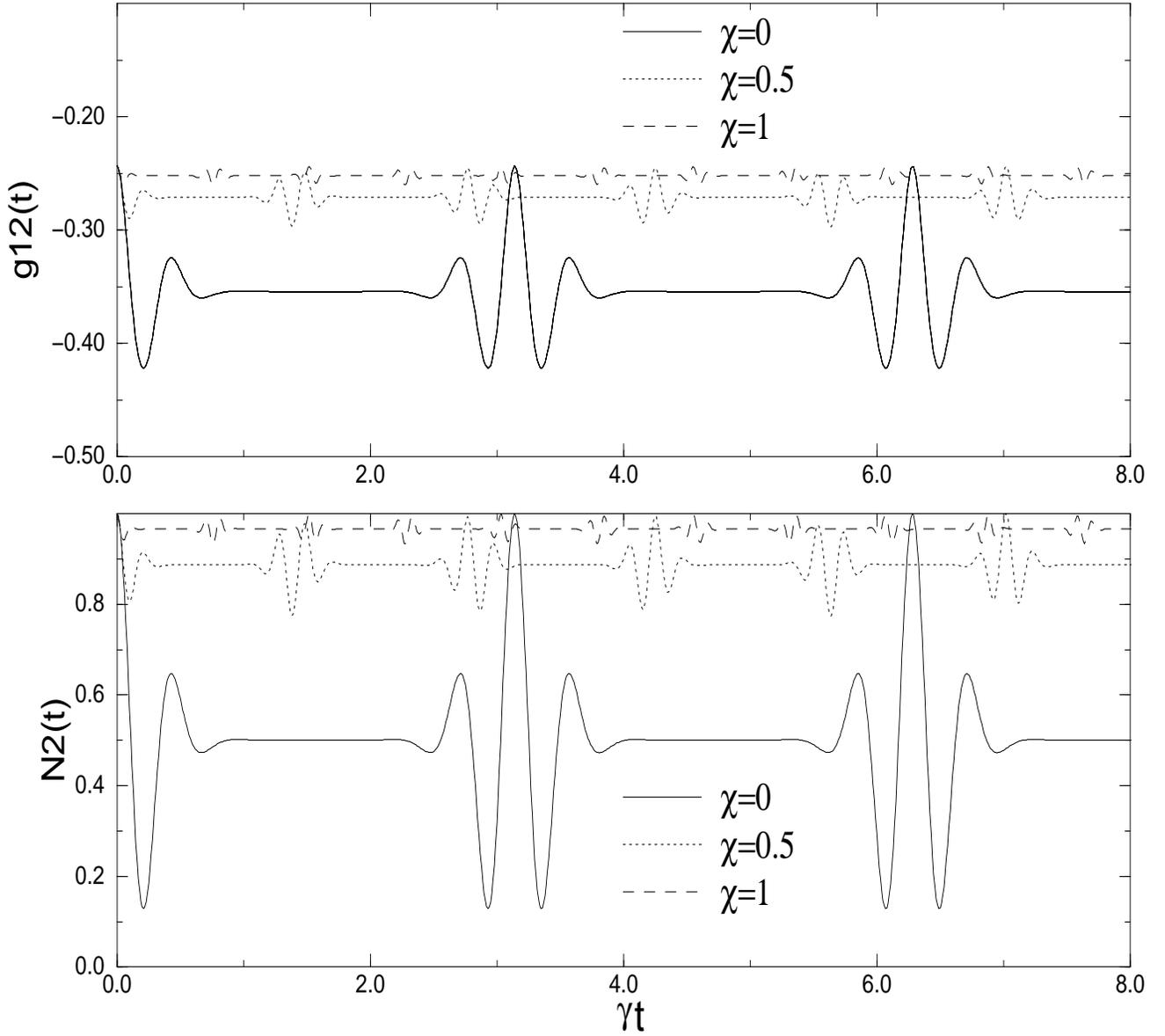} 
\caption{Figure 3.  Same as figure 1 but field
in pair coherent field.}
\label{fig3}
\end{figure}
\end{center}
\end{document}